\documentclass{jltp}
\usepackage{graphicx}
\usepackage{bm}

\title{Matter wave solitons at finite temperatures}

\author{B.~Jackson, C.~F.~Barenghi, and N.~P.~Proukakis}

\address{School of Mathematics and Statistics, University of Newcastle
upon Tyne, \\ NE1 7RU, United Kingdom}

\runninghead{B.~Jackson, C.~F.~Barenghi, and N.~P.~Proukakis}
{Matter wave solitons at finite temperatures}

\begin{document}

\maketitle

\begin{abstract}

We consider the dynamics of a dark soliton in an elongated harmonically 
trapped Bose-Einstein condensate. A central question concerns the behavior 
at finite temperatures, where dissipation arises due to
the presence of a thermal cloud. We study 
this problem using coupled Gross-Pitaevskii and $N$-body simulations, which 
include the mean field coupling between the condensate and thermal cloud. We 
find that the soliton decays relatively quickly even at very low temperatures,
with the decay rate increasing with rising temperature.

PACS numbers: 03.75.Lm, 05.45.Yv, 67.80.Gb
\end{abstract}

\maketitle

\section{INTRODUCTION}

Dark solitons represent a ubiquitous feature of nonlinear 
systems\cite{kivshar98}, and their experimental realization in atomic 
Bose-Einstein
condensates (BECs)\cite{burger99,denschlag01,dutton01,anderson01} has opened 
up new possibilities for their study. The stability of the soliton, and its
dependence upon the geometry of the condensate, is a problem of considerable
interest. In three dimensions a soliton is dynamically unstable to 
undulations in its transverse profile. This so-called 
``snake'' instability leads to the soliton decaying to vortex lines or rings,
and has been observed experimentally\cite{dutton01,anderson01}. This 
instability, however, can be suppressed by confining the condensate much more 
strongly in the transverse direction than longitudinally, so that the 
condensate is highly elongated along the axial direction.

A second decay mechanism arises from the fact that at non-zero temperatures 
the condensate coexists with a noncondensed cloud composed of thermally 
excited quasiparticles. Interactions of the soliton with the thermal cloud
leads to dissipation, so that the soliton loses energy and accelerates.
Experiments\cite{burger99} suggest that the timescale for this decay can be 
relatively
short, on the order of 10\,{\rm ms}, so has important consequences for the 
dynamics. This decay process has previously been studied
theoretically by Fedichev {\it et al.}\cite{fedichev99} and Muryshev 
{\it et al.}\cite{muryshev02} by considering the reflection of thermal 
excitations from the soliton. The role of quantum fluctuations was addressed
in Dziarmaga {\it et al.}\cite{dziarmaga03}.

In this paper we describe the use of numerical 
simulations\cite{jackson02} to model a soliton at finite temperatures in 
a highly elongated condensate, where the trap frequency in the radial 
direction, $\omega_{\perp}$, greatly exceeds that along the axial direction, 
$\omega_z$. At $T=0$ the soliton is 
predicted\cite{busch00,frantzeskakis02}, on the basis 
of the Gross-Pitaevskii (GP) equation, to oscillate longitudinally at a 
frequency of 
$\omega_z / \sqrt{2}$, which involves continuous undamped soliton-sound
interactions\cite{parker03}. We follow the 
motion of the soliton as a function of time, finding the expected 
oscillation along the axial direction. At finite temperature the
energy loss of the soliton due to the presence of the thermal cloud leads to 
an increase in the amplitude of the oscillation.  
We investigate this soliton decay process as a function
of temperature, and find that it is appreciable even at very low $T$. 
 
\section{THEORY}

Our treatment of finite temperatures is based on the ZNG 
formalism\cite{zaremba99}, where we solve the following coupled equations 
\begin{equation}
 i \hbar \frac{\partial \Psi}{\partial t} = \left (-\frac{\hbar^2 \nabla^2}
 {2m} + V + g n_c + 2g \tilde{n} \right) \Psi, 
\label{eq:GP}
\end{equation}
\begin{equation}
 \frac{\partial f}{\partial t} +\frac{{\bm p}}{m} \cdot {\bm \nabla} f - 
 {\bm \nabla} U \cdot {\bm \nabla}_{\bm p} f =0. 
\label{eq:Boltz}
\end{equation}
The former is a generalized GP equation for the 
condensate
wavefunction $\Psi({\bm r}, t)$, and the latter is a Boltzmann equation for
the thermal cloud phase space density $f({\bm p},{\bm r},t)$. Here the 
condensate and thermal cloud densities are defined as $n_c = |\Psi|^2$ and 
$\tilde{n}=\int f \,d{\bm p}/h^3$ respectively, while
$g=4\pi \hbar^2 a/m$ parametrizes the mean field interactions between atoms of 
mass $m$ with scattering length $a$. The effective potential is 
$U=V+2g(n_c+\tilde{n})$, with $V=m(\omega_{\perp}^2 r^2+\omega_z^2 z^2)/2$ 
representing the harmonic trap confining the atoms.
Note that for the purposes of this paper we neglect collisions between the 
atoms, and the coupling between the condensate and thermal cloud is purely
mean field in nature. For the parameters considered here, inclusion of 
collisions is
expected to slightly increase the damping rate, but this will be considered 
explicitly in future work.

The initial condition for our simulations is found by self-consistently
solving (\ref{eq:GP}) in imaginary time to find the ground state wavefunction,
which is coupled to a thermal cloud described by a Bose distribution 
$f({\bm p},{\bm r})=\{ \exp[(p^2/2m+U-\mu)/k_B T]-1 \}^{-1}$, where $\mu$ is 
the condensate chemical potential.
The methods for solving Eqs.\ (\ref{eq:GP}) and (\ref{eq:Boltz}) for the 
subsequent finite temperature dynamics are described by Jackson and 
Zaremba\cite{jackson02}. A difference with respect to this earlier work is 
that here the GP equation (\ref{eq:GP}) is solved using cylindrical 
coordinates with a Crank-Nicholson time-stepping scheme. Since the system is
cylindrically symmetric the problem effectively reduces to two dimensions,
improving the efficiency of the simulations.

\section{RESULTS}

Finite temperature simulations have been performed for $N=2\times 10^4$ 
$^{87}{\rm Rb}$ atoms in a trap with frequencies 
$\omega_{\perp}=2\pi\times 2500\,{\rm Hz}$ and 
$\omega_z = 2\pi\times 10\,{\rm Hz}$, resulting in 
a cloud that is highly elongated along the axial direction. The critical
temperature for Bose condensation is close to that for
an ideal gas in the thermodynamic limit\cite{pitaevskii}, $T_c^0 = 0.941\hbar 
(\omega_{\perp}^2 \omega_z N)^{1/3}/k_B$, which 
for these parameters corresponds to $T_c^0=486\,{\rm nK}$. At time $t=0$ the 
equilibrium configuration for a given temperature is imprinted with a
dark soliton, by multiplying the condensate wavefunction $\Psi ({\bm r})$ 
with\cite{parker03} $\Psi_s (z) = \beta \tanh (\beta z/\xi)
 + i v/c$,
where $\beta=\sqrt{1-(v/c)^2}$, $\xi=\hbar/\sqrt{mgn}$ is the condensate 
healing length, and $c=\sqrt{gn/(2m)}$ is the speed of sound for density $n$. 
For the 
benefit of the following discussion, it is useful to note that for a condensate
of uniform density $n_0$ the soliton energy is given by\cite{kivshar98}
$E_s = 4\hbar c n_0 \beta^3/3$, while the 
density at the soliton minimum (equal to the difference between the 
background density and the depth of the soliton) is\cite{pitaevskii} 
$n_s=n_0 v^2/c^2$. So an increase in $v$ corresponds to a 
decrease in the soliton energy and depth, which both tend to zero as $v \to c$.

\begin{figure}[h]
\centering \scalebox{0.69}
 {\includegraphics{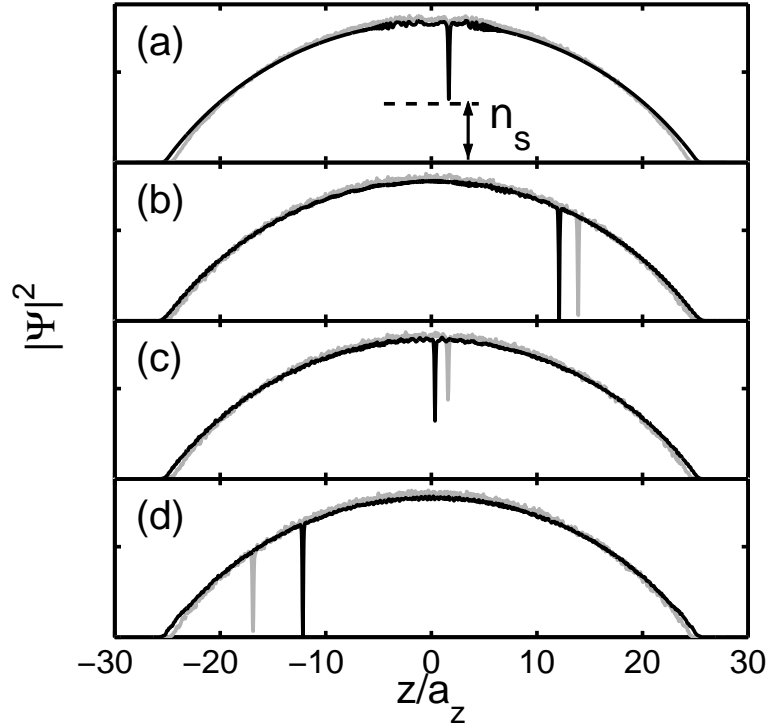}}
\caption{Density $|\Psi|^2$ along an axial cross-section of the 
 condensate, showing the evolution of a dark soliton with initial velocity
 $v=0.5c$. The black lines plot the densities for $T=0$, while the gray lines
 are for $T=150\,{\rm nK}$, at times of (a) $\omega_z t=0.2$, 
 (b) $\omega_z t=2.2$, (c) $\omega_z t = 4.4$, and (d) $\omega_z t = 6.7$. The
 axial position $z$ is in units of $a_z = \sqrt{\hbar/(m\omega_z)}$, while 
 the density is in arbitrary units. The density at the soliton minimum, $n_s$, 
 is labelled in (a).}
\label{fig:sol-xs}
\end{figure}

\begin{figure}[h]
\centering \scalebox{0.69}
 {\includegraphics{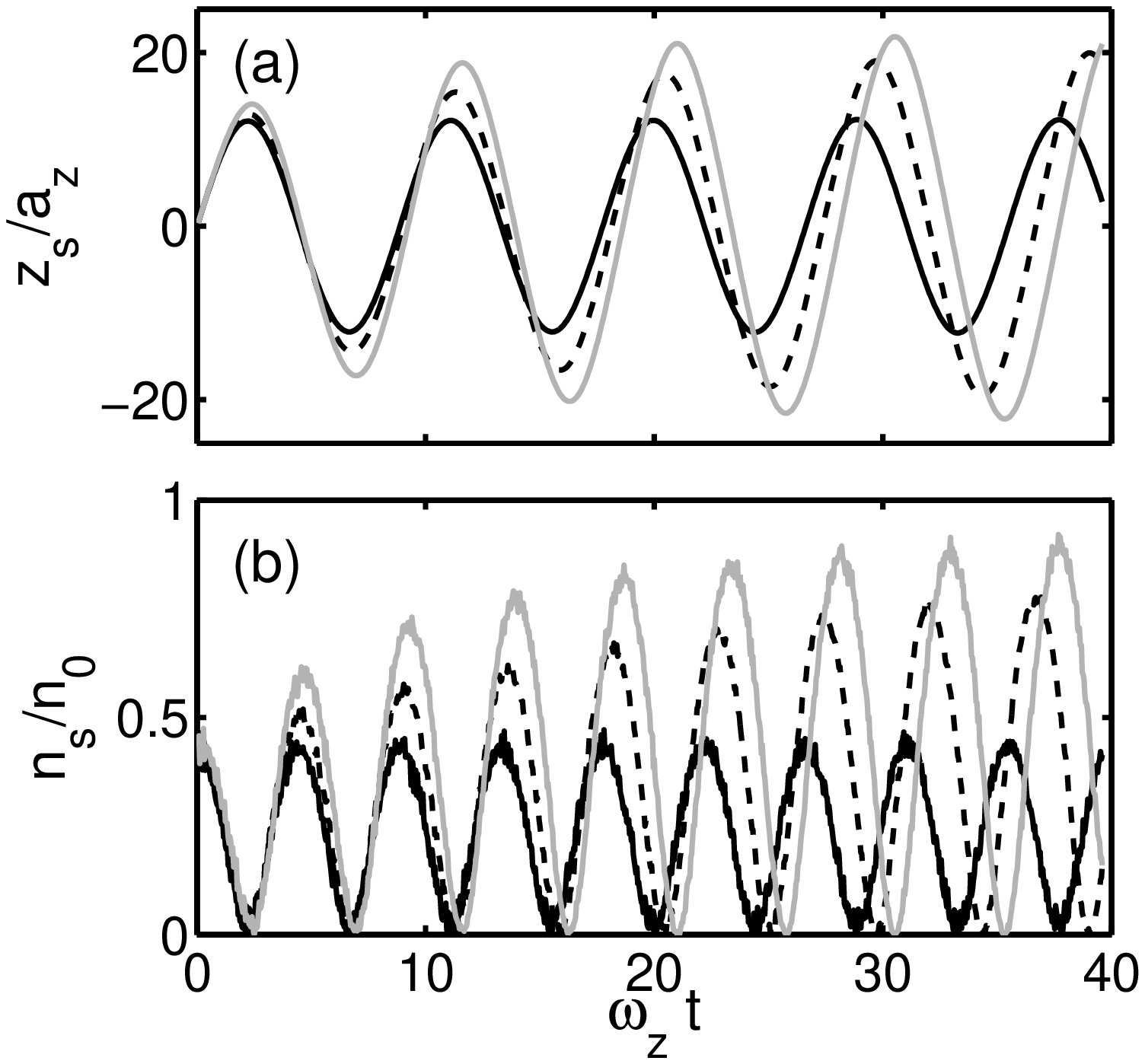}}
\caption{(a) Axial position (in units of $a_z$) of the soliton minimum as a 
 function of time, where initially $v=0.5c$. The curves show results for
 different temperatures, with $T=0$ (solid black), $T=100\, {\rm nK}$ 
 (dashed), and $T=150\, {\rm nK}$ (gray). (b) Density at the soliton minimum
 $n_s$ (divided by the central density of the condensate without a soliton, 
 $n_0$) as a function of time, where the curves are labelled as in (a).}
\label{fig:solmin}
\end{figure}

For $T=0$ the dynamics are simply given by the 
GP equation (\ref{eq:GP}) with $\tilde{n}=0$. The black lines 
in Fig.~\ref{fig:sol-xs} plot the subsequent evolution of the density at early
times, where the soliton oscillates along the axial direction. 
The position
of the soliton minimum, $z_s$, is plotted in Fig.~\ref{fig:solmin} (a), 
illustrating
that the soliton oscillates at constant amplitude with a frequency of 
$\omega_z/\sqrt{2}$, as expected\cite{busch00,frantzeskakis02,parker03}. 
The parameter $n_s$, representing the density at the soliton minimum as 
labelled in Fig.~\ref{fig:sol-xs} (a), acquires its
maximum value at the center of the trap and is zero at the turning points
of the soliton motion. Fig.~\ref{fig:solmin} (b) plots $n_s$, divided by the central 
density of the condensate without a soliton, $n_0$. The peak value of this
parameter (attained at the trap centre) remains constant, as expected for 
undamped motion.

The corresponding density cross-sections for $T=150\,{\rm nK}$ are plotted 
with gray lines  
in Fig.~\ref{fig:sol-xs}, where one sees that, even during the first 
oscillation, the presence of the thermal cloud
leads to an increase in the amplitude of the oscillation 
(Fig.~\ref{fig:sol-xs} (b) and (d)), in addition to a
significant decrease in the depth of the soliton when it is at 
the trap center (Fig.~\ref{fig:sol-xs} (c)).
The steady rise in oscillation amplitude is also shown in 
Fig.~\ref{fig:solmin} (a),
while the decrease in soliton depth at $z_s=0$ (corresponding to an increase 
in $n_s$) is illustrated in 
Fig.~\ref{fig:solmin} (b). This accompanies an increase in velocity at 
$z_s=0$, and reflects an energy decay as 
the soliton interacts with the thermal cloud. 

Simulations have also been performed at other temperatures, and one finds 
less damping at lower $T$ (as illustrated for $T=100\,{\rm nK}$ in 
Fig.~\ref{fig:solmin}) and more damping at higher $T$, as expected.
However, it is remarkable that appreciable damping occurs
even for these relatively low temperatures ($T<0.3 T_c$) , which contrasts 
with the situation for collective modes where
the effects of finite temperature tend to be significant only at higher 
temperatures\cite{jin97,stamperkurn98,marago01}. Another noteworthy feature 
relates to the time dependence of the amplitude, the increase of which
appears to saturate as it
approaches the radius of the condensate. This effect has also been observed 
for $v=0.25c$, where the initial oscillation amplitude is smaller, and hence 
the saturation does not occur until later times.
 
\section{CONCLUSIONS}

We have performed numerical simulations to study the dynamics of dark solitons
in a Bose-Einstein condensate at different temperatures. Our numerics 
for zero temperature reproduce  
soliton oscillations along the axial direction at constant amplitude and at 
a frequency of $\omega_z/\sqrt{2}$. However, at finite temperature, 
mean field coupling between the condensate and thermal cloud is shown to 
lead to a lower frequency
oscillation with a steadily increasing amplitude. This corresponds to the 
soliton losing energy to the thermal cloud, with a rate that increases with 
rising initial temperature. 

Note that this energy transfer cannot be considered as a heating of the 
thermal cloud, since temperature is not defined during the simulation due 
to a lack of rethermalizing collisions. Future calculations will study the 
influence of these collisional processes, as well
as modelling the soliton decay for the experimental parameters of Burger
{\it et al.}\cite{burger99}. 
A further motivation for future work is to ascertain the feasibility of 
Proukakis {\it et al.}\cite{proukakis04}, where a parametric driving
scheme was proposed for stabilization of a soliton against a 
phenomenologically modelled thermal cloud.
We plan to revisit this problem with
our more detailed finite temperature simulations. 

\section*{ACKNOWLEDGMENTS}

This research is supported by the UK Engineering and
Physical Sciences Research Council. We thank N.~G.~Parker for useful 
discussions.


\begin{thebibliography}{99}

\bibitem{kivshar98}
 Y.~S.~Kivshar and B.~Luther-Davies, {\it Phys. Rep.} {\bf 298}, 81 (1998).

\bibitem{burger99} 
 S.~Burger {\it et al.}, {\it Phys. Rev. Lett.} {\bf 83}, 5198 (1999). 

\bibitem{denschlag01}
 J.~Denschlag {\it et al.}, {\it Science} {\bf 287}, 97 (2001).

\bibitem{dutton01}
 Z.~Dutton {\it et al.}, {\it Science} {\bf 293}, 663 
 (2001).

\bibitem{anderson01}
 B.~P.~Anderson {\it et al.}, {\it Phys. Rev. Lett.} {\bf 86}, 2926 (2001).

\bibitem{fedichev99}
 P.~O.~Fedichev, A.~E.~Muryshev, and G.~V.~Shlyapnikov, {\it Phys. Rev. A} 
 {\bf 60}, 3220 (1999).

\bibitem{muryshev02}
 A.~E.~Muryshev {\it et al.}, {\it Phys. Rev. Lett.} {\bf 89}, 110401 (2002).

\bibitem{dziarmaga03}
 J.~Dziarmaga, Z.~P.~Karkuszewski and K.~Sacha, {\it J.~Phys.~B}, 36
 1217 (2003). 

\bibitem{jackson02}
 B.~Jackson and E.~Zaremba, {\it Phys.~Rev.~A} {\bf 66}, 033606 (2002).

\bibitem{busch00}
 Th.~Busch and J.~R.~Anglin, {\it Phys.~Rev.~Lett.} {\bf 84}, 2298 (2000).

\bibitem{frantzeskakis02} 
 D.~J.~Frantzeskakis {\it et al.}, {\it Phys.~Rev.~A} {\bf 66}, 053608 (2002).

\bibitem{zaremba99}
 E.~Zaremba, T.~Nikuni, and A.~Griffin, {\it J.~Low.~Temp.~Phys.} {\bf 116}, 
 277 (1999).

\bibitem{parker03}
 N.~G.~Parker {\it et al.}, {\it Phys.~Rev.~Lett.} {\bf 90}, 220401 (2003).

\bibitem{pitaevskii}
 L.~P.~Pitaevskii and S.~Stringari, {\it Bose-Einstein Condensation},
 Clarendon Press, Oxford (2003).

\bibitem{jin97}
 D.~S.~Jin {\it et al.}, {\it Phys.~Rev.~Lett.} {\bf 78}, 764 (1997). 

\bibitem{stamperkurn98} 
 D.~M.~Stamper-Kurn {\it et al.}, {\it Phys.~Rev.~Lett.} {\bf 81}, 500 (1998).
 
\bibitem{marago01} 
 O.~M.~Marag\'{o} {\it et al.}, {\it
 Phys.~Rev.~Lett.} {\bf 86}, 3938 (2001).

\bibitem{proukakis04} 
 N.~P.~Proukakis {\it et al.}, {\it Phys.~Rev.~Lett.} {\bf 93}, 130408 (2004).

\end{thebibliography}
\end{document}